\UseRawInputEncoding
\documentclass[%
 reprint,
 amsmath,amssymb,
 aps,
]{revtex4-2}

\usepackage{graphicx}
\usepackage{dcolumn}
\usepackage{bm}

\begin{document}

\preprint{APS/123-QED}

\title{Gain-gain and gain-lossless PT-symmetry broken from PT-phase diagram}

\author{Qi Zhang,$^{1,2}$ Yun Ma,$^{1}$ Qi Liu,$^{1,3}$ Xinchen Zhang,$^{1}$ Yali Jia,$^{1}$ \\ LiMin Tong,$^{6,7,8}$ Qihuang Gong,$^{1,3,4,5}$ and Ying Gu$^{1,3,4,5}$}

 \altaffiliation{ygu@pku.edu.cn}

  \affiliation{%
 $^{1}$State Key Laboratory for Mesoscopic Physics, Department of Physics, Peking University, Beijing 100871, China\\
 $^{2}$Institute of Navigation and Control Technology, China North Industries Group Corporation, Beijing 100089, China\\
$^{3}$Frontiers Science Center for Nano-optoelectronics \& Collaborative Innovation Center of Quantum Matter \& Beijing Academy of Quantum Information Sciences, Peking University, Beijing 100871, China\\
$^{4}$Collaborative Innovation Center of Extreme Optics, Shanxi University, Taiyuan 030006, China\\
$^{5}$Peking University Yangtze Delta Institute of Optoelectronics, Nantong 226010, China\\
$^{6}$Interdisciplinary Center for Quantum Information, State Key Laboratory of Modern Optical Instrumentation, College of Optical Science and Engineering, Zhejiang University, Hangzhou 310027, China\\
$^{7}$Jiaxing Key Laboratory of Photonic Sensing \& Intelligent Imaging, Jiaxing 314000, China\\
$^{8}$Intelligent Optics \& Photonics Research Center, Jiaxing Research Institute Zhejiang University, Jiaxing 314000, China\\
}%

\date{\today}

\begin{abstract}
Parity–time (PT) symmetry and broken in micro/nano photonic structures have been investigated extensively as they bring new opportunities to control the flow of light based on non-Hermitian optics. Previous studies have focused on the situations of PT-symmetry broken in loss-loss or gain-loss coupling systems. Here, we theoretically predict the gain-gain and gain-lossless PT-broken from phase diagram, where the boundaries between PT-symmetry and PT-broken can be clearly defined in the full-parameter space including gain, lossless and loss. For specific micro/nano photonic structures, such as coupled waveguides, we give the transmission matrices of each phase space, which can be used for beam splitting. Taking coupled waveguides as an example, we obtain periodic energy exchange in PT-symmetry phase and exponential gain or loss in PT-broken phase, which are consistent with the phase diagram. The scenario giving a full view of PT-symmetry or broken, will not only deepen the understanding of fundamental physics, but also will promote the breakthrough of photonic applications like optical routers and beam splitters.
\end{abstract}

\maketitle


\section{Introduction}

The investigation of parity-time (PT) symmetry extends the framework of quantum mechanics into complex domain and shows novel physical properties \cite{2019Exceptional}. In 1998, Bender and Boettcher \cite{1998Real} first put forward that as long as that parity-time symmetry is satisfied, even non-Hermitian Hamiltonians can exhibit entirely real eigenvalue spectra. The most interesting effect related to this non-Hermitian Hamiltonian is the phase transition behavior arising from a spontaneous breakdown of parity and time symmetry \cite{1998PT,2004Physics,2002Complex}. In many fields of theoretical physics, PT symmetry is developed rapidly, such as quantum field theories \cite{2001Bound}, complex crystals \cite{PhysRevA.64.042716,2001PT} and Lie algebras \cite{2000sl}. Recently, it has been recognized that PT symmetry can also be experimentally explored and ultimately used in optics \cite{2017Non,2018Non,2019Exceptional}. When gain and loss are introduced into the optical systems, the appearance of exceptional points (EPs) greatly changes the response of the system: abrupt phase transitions occurs, leading to a series of unique properties like nonreciprocal transmission \cite{2010Unidirectional,2014Parity,2014Parity1}, unidirectional invisibility \cite{2010Unidirectional,AosRnsurr2012Parity,2013Experimental}, loss-induced transparency \cite{PhysRevLett.103.093902}, and chiral modes \cite{PhysRevA.89.012119,Kim:14}.

In micro/nano photonic structures, early studies about PT symmetry began in two coupled waveguides with dielectric constants of complex conjugate. In theory, researchers took the EP to divide PT-symmetry and PT-broken by solving coupled mode equations, which can be tuned by changing coupling coefficient and gain or loss rate \cite{2010Observation,AosRnsurr2012Parity}. Considering the difficulty of preparing gain medium, loss-only micro/nano waveguide structures are usually used to construct equivalent optical potential experimentally \cite{PhysRevLett.112.143903,PhysRevLett.103.093902,Ornigotti_2014}. Moreover, going beyond two waveguides, PT symmetry can be also observed in whispering gallery microresonators. In 2014, non-reciprocal light transmission was realized in the PT-broken phase in active-passive-coupled microdisks \cite{2014Parity}. With the reduction of resonances of two whispering gallery modes, spontaneous PT symmetry breaking occurs at fixed gain-to-loss ratio is also found \cite{2014Parity}. Utilizing the unique phase transition properties of EP formed by whispering gallery modes, loss-induced suppression and revival of Raman lasing \cite{doi:10.1126/science.1258004}, mode suppression \cite{doi:10.1126/science.1258479} and single mode lasing \cite{doi:10.1126/science.1258479,doi:10.1126/science.1258480} are observed in microresonators. Although PT symmetry in micro/nano photonic structures has been studied a lot, two aspects still worth to be investigated. Such as there is no full-parameter space phase diagram containing loss, lossless and gain to describe PT-symmetry and PT-broken. Besides, the PT symmetry properties in gain-gain and gain-lossless micro/nano structures have not been studied.

Beam splitting plays an important role in the basic researches and applications of optical physics, and its properties are generally described in the form of transmission matrices \cite{Christian1999Atomic,2007Analytical}. Depending on the nature of the input and output light, beam splitting can be used in quantum and classical optics \cite{doi:10.1119/1.12387}. Among them, classical beam splitting can be used to design optical routers \cite{2012All}, logic gates \cite{ZHANG1992185, Optical_waveguides} and other devices \cite{doi:10.1063/1.3279134, Subwavelength}, while quantum beam splitting is valuable in studying anti-bunching \cite{PhysRevA.57.2134}, squeezed states \cite{PhysRevLett.39.691} and quantum entanglement \cite{PhysRevLett.59.2044, connect}. However, most of previous studies on beam splitting have focused on the Hermitian system \cite{doi:10.1080/09500340008232434}. Using PT symmetry or broken in non-Hermitian systems to manipulate light propagation has rarely been studied.

Here, we first drew the full-parameter space phase diagram containing gain, lossless and loss. Two EP-lines divide the parameter space into two regions, called PT-symmetry and PT-broken. It can be used to predict the phase of coupled systems. For coupled waveguides, we derived the transmission matrices of each phase space and obtained the PT-broken conditions, which can improve the flexibility of beam splitting design. Several specific cases, especially the gain-gain and gain-lossless coupled waveguides, were selected for numerical calculation. We observed periodic energy exchange in PT-symmetry phase and exponential gain or loss in PT-broken phase. The results are consistent with the predictions from the phase diagram. The above contents are expected to be applied to quantum beam splitting, quantum state preparation and other quantum information processes, which will provide guidance for basic physics researches and promote the development of micro/nano photonic applications.

\section{PT Phase Diagram in Full-parameter Space}

\begin{figure*}
\includegraphics[width=12cm]{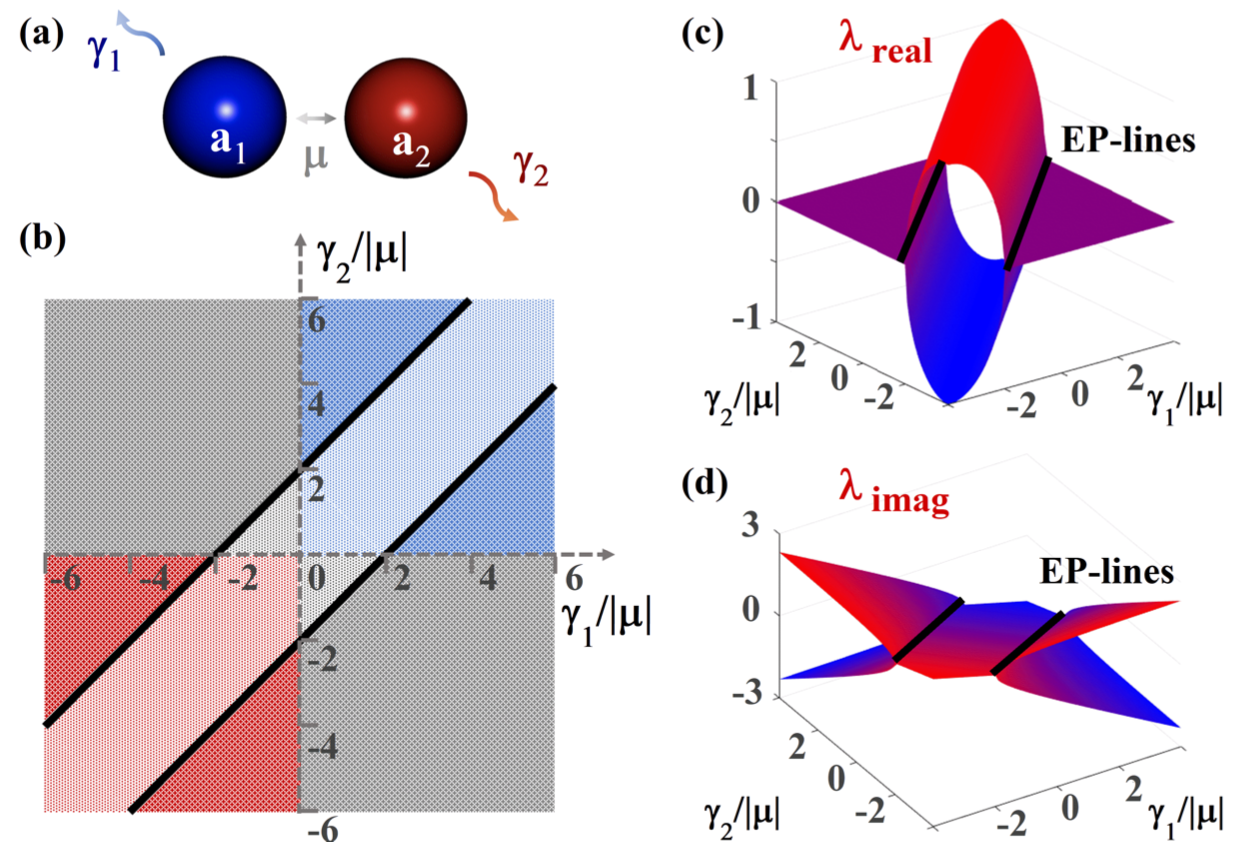}
\caption{\label{fig:wide}(a) Generic system of two coupled modes. The blue ($a_{1}$) and red ($a_{2}$) circles represent two modes. $\mu$ is the coupling coefficient, and $\gamma_{1}$, $\gamma_{2}$ are the decay rates ($\gamma_{1},\gamma_{2}>0$) or gain rates ($\gamma_{1},\gamma_{2}<0$) respectively. (b) Two-dimensional phase diagram in parameter space $(\frac{\gamma_{1}}{|\mu|}, \frac{\gamma_{2}}{|\mu|})$. The red, blue and grey areas are gain-gain, loss-loss and gain-loss coupling respectively. Two EP-lines (black diagonal lines) divide the space into PT-symmetry phase and PT-broken phase. The real (c) and imaginary (d) parts of the eigenvalues $\lambda$ varying with $\frac{\gamma_{1}}{|\mu|}$ and $\frac{\gamma_{2}}{|\mu|}$. In PT-symmetry phase, $\lambda_{real}$ separated and $\lambda_{imag}$ degenerated. On the contrary, in PT-broken phase, $\lambda_{real}$ degenerated and $\lambda_{imag}$ separated.}
\end{figure*}

Precisely manipulation of the transmission of light could dramatically facilitate the performance of quantum devices. To develop on-chip quantum devices, the combination of PT symmetry and micro/nano photonics holds great promise. In this part, we drew the PT phase diagram of two photonic structures in the full-parameter space including gain, lossless and loss. Using this phase diagram, we predicted the condition of PT-symmetry and PT-broken phase. In previous studies, there are many researches on loss-loss or gain-loss modes coupling systems \cite{AosRnsurr2012Parity}, but almost no study on gain-gain or gain-lossless systems. In the next sections, we drove transmission matrices and did numerical calculations of coupled waveguides based on PT phase diagram.

As shown in Fig. 1(a), two closely spaced general modes (represented by blue and red circles) in a coupled system exchange energy with coupling coefficient $\mu$ and gain or loss rates $\gamma_1$, $\gamma_2$ respectively, where the positive number ($\gamma>0$) for loss and the negative number ($\gamma<0$) for gain. Assuming that $a_{1}(\xi)$ and $a_{2}(\xi)$ represent the amplitude evolution of the two modes with variable $\xi$ respectively, where $\xi$ can mean either propagating distance or evolution time. According to the coupled mode theory \cite{Huang:94},
\begin{equation}
i \frac{d}{d \xi}\left(\begin{array}{c}
a_{1}(\xi) \\
a_{2}(\xi)
\end{array}\right)=-\left(\begin{array}{cc}
i \gamma_{1} & \mu \\
\mu^{*} & i \gamma_{2}
\end{array}\right)\left(\begin{array}{c}
a_{1}(\xi) \\
a_{2}(\xi)
\end{array}\right).
\end{equation}
In combination with the above two differential equations in Eq. (1), one variable can be eliminated that
\begin{equation}
    \frac{d^{2} a_{1}(\xi)}{d \xi^{2}}+\left(\gamma_{1}+\gamma_{2}\right) \frac{d a_{1}(\xi)}{d \xi}+\left(\gamma_{1} \gamma_{2}+|\mu|^{2}\right) a_{1}(\xi)=0.
\end{equation}
The eigenvalues of the exponential term in the general solutions are
\begin{equation}
    \lambda_{\pm}=\frac{1}{2}i\left(\gamma_{1}+\gamma_{2}\right) \mp \frac{1}{2} i\sqrt{\left(\gamma_{1}-\gamma_{2}\right)^{2}-4|\mu|^{2}}.
\end{equation}
According to the form of the second part $\sqrt{\left(\gamma_{1}-\gamma_{2}\right)^{2}-4|\mu|^{2}}$, the photonic system can be divided into two regions: PT-symmetry phase and PT-broken phase, where the non-Hermitian properties are quite the opposite. Varying the parameters $\mu$, $\gamma_1$ or $\gamma_2$, two eigenvalues as well as eigenmodes will coincide at
\begin{equation}
    \frac{\gamma_{1}}{|\mu|}-\frac{\gamma_{2}}{|\mu|}=\pm 2,
\end{equation}
which form EP-lines. It indicates that PT symmetry broken can occur not only in the gain-loss system, but also in other systems \cite{PhysRevLett.112.143903,PhysRevLett.103.093902}. Among them, gain-gain coupling and gain-lossless coupling are rarely involved. In PT-broken phase, the form of transmission matrix is completely different from the usual case, which is expected to provide a new degree of freedom for the control of quantum states \cite{2019Exceptional}.

We drew the PT phase diagram in the full-parameter space as Fig. 1(b). Taking the values of independent variables $\frac{\gamma_{1}}{|\mu|}$ and $\frac{\gamma_{2}}{|\mu|}$ as two coordinate axes, the whole space can be divided into four quadrants: the blue area for loss-loss coupling in the first quadrant, the red area for gain-gain coupling in the third quadrant, and the grey area for gain-loss coupling in the second or the fourth quadrant. The sparse and dense shadow express PT-symmetry and PT-broken phase of corresponding areas. The central point indicates that neither mode has gain or loss, only energy exchange exists between modes, and the remaining positions on the axis represent a lossless mode coupled to a gain or loss mode. In previous studies of PT symmetry, the systems with coupled gain and loss modes attracted the most attention (grey area) \cite{AosRnsurr2012Parity,2010Observation}. In recent years, two different loss modes or loss-lossless coupling systems have also been used to achieve PT-broken experimentally (blue area) \cite{PhysRevLett.112.143903,PhysRevLett.103.093902}. However, there were few studies on gain-gain or gain-lossless coupling systems, whose setup parameters lie in red area. Thus the phase diagram we proposed can be used to predict whether these systems respect to PT-symmetry or not.

The two black oblique lines in Fig. 1(b) are $\frac{\gamma_{1}}{|\mu|} - \frac{\gamma_{2}}{|\mu|} = \pm 2 $, corresponding to the degeneracy of the eigenvalues in Eq. (3), which called EP-lines. In the region between two EP-lines in the phase diagram (sparse shadow of corresponding colors), the eigenvalues are complex, which means system supports oscillating modes, and eigenmodes satisfy PT-symmetry. Yet on the top left and bottom right areas of the EP-lines (dense shadow of corresponding colors), the system reaches PT-broken phase, showing exponential gain or loss. For gain-gain coupling in the third quadrant, to reach PT-broken phase, the coupling coefficient and the gain rates have to satisfy the condition $|\gamma_{1}-\gamma_{2}|>2|\mu|$. While for gain-lossless coupling, $\gamma_{1}(\gamma_{2})=0$, the PT-broken condition becomes $\gamma_{2}(\gamma_{1})>2|\mu|$. Different regions in the phase diagram correspond to different forms of eigenvalues and mode evolution. Fig. 1(c) and (d) show the real and imaginary parts of the eigenvalues in two-dimensional parameters space $(\frac{\gamma_{1}}{|\mu|}, \frac{\gamma_{2}}{|\mu|})$ that described by Eq. (3). Two black lines in each figure express EP-lines that $\frac{\gamma_{1}}{|\mu|}-\frac{\gamma_{2}}{|\mu|} = \pm 2$, which divide the plane into two distinct phase spaces, PT-symmetry (the area between EP-lines) and PT-broken (the area on either side of the line). On EP-lines, both real and imaginary parts of the eigenvalues coalesce. The regions where the real parts separate in Fig. 1(c) correspond to PT-symmetry phase of the system, while the regions where the real parts degenerate correspond to the PT-broken phase of the system.

If $\lambda=\lambda_{real}+i\lambda_{imag}$, in the PT-symmetry phase, the term $\left(\gamma_{1}-\gamma_{2}\right)^{2}-4|\mu|^{2}$ is negative and the eigenvalues can be expressed as
\begin{equation}
    \left\{\begin{array}{c}
\lambda_{\text {real }}=\pm \frac{|\mu|}{2} \sqrt{4-\left(\frac{\gamma_{1}}{|\mu|}-\frac{\gamma_{2}}{|\mu|}\right)^{2}} \\
\lambda_{\text {imag }}=\frac{|\mu|}{2}\left(\frac{\gamma_{1}}{|\mu|}+\frac{\gamma_{2}}{|\mu|}\right) 
\end{array}\right. .
\end{equation}
For the real part, two eigenvalues separate with opposite sign of each other, corresponding to the region between EP-lines in Fig. 1(c). For the imaginary part, two eigenvalues are degenerate and the values depend on the magnitude of $\frac{\gamma_{1}}{|\mu|}$, $\frac{\gamma_{2}}{|\mu|}$, corresponding to the inclined plane in the central of  Fig. 1(d). Hence, in the PT-symmetry phase the evolution of the two eigenmodes are dominated by oscillation and accompanied by loss or gain. 

On the other hand, in the PT-broken phase, the eigenvalues become pure imaginary numbers and are written as
\begin{equation}
    \left\{\begin{array}{c}
\lambda_{\text {real }}=0 \\
\lambda_{\text {imag }}=\frac{|\mu|}{2}\left[(\frac{\gamma_{1}}{|\mu|}+\frac{\gamma_{2}}{|\mu|}) \mp \sqrt{\left(\frac{\gamma_{1}}{|\mu|}-\frac{\gamma_{2}}{|\mu|}\right)^{2}-4}\right] 
\end{array}\right. .
\end{equation}
The real part $\lambda_{real}$ is identical to 0, corresponding to the plane of the edge in Fig. 1(c). The imaginary part $\lambda_{imag}$ is separated and represented as a symmetrical surface on both sides of the EP-lines in Fig. 1(d). The symmetry of each mode is broken such that one of them enjoys amplification and the other one experiences attenuation. Numerical analysis of eigenvalues is helpful for further understanding of their characteristics, and lays a foundation for utilizing their unique mode evolution in different phase spaces.

The investigation of mode dynamics in PT symmetry system will be of great benefit to the study of non-Hermitian Hamiltonian in the field of micro/nano photonics. The evolution characteristics of modes in different photonic systems are quite different. For example, in coupled waveguide system, PT-symmetry corresponds to oscillation modes, while PT-broken corresponds to exponential gain or decay modes \cite{O2019Parity,ElGanainy07}. Not limited to the waveguide coupling structures mentioned above, a range of different photonics systems that governed by the coupled mode equations can be described, for example, coupled optical cavities \cite{PhysRevLett.108.173901,0Enhanced}, counter-propagating waves \cite{PhysRevA.86.033801} and coupled orthogonal polarization states \cite{PhysRevLett.118.093002}, etc. By means of the full-parameter space phase diagram we drew, PT broken conditions of gain-gain and gain-lossless coupling systems can be predicted, which have not been studied.

\section{Beam Splitting and Transmission Matrices}

\begin{figure}[ht!]
\centering\includegraphics[width=7cm]{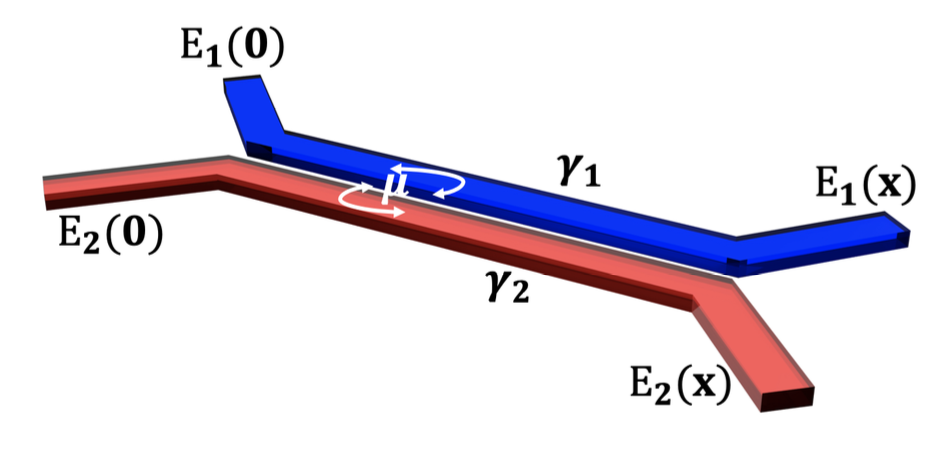}
\caption{Optical beam splitting composed of two coupled waveguides. $E_{1}(x)$ and $E_{2}(x)$ are the electric field intensities, $\mu$ is the coupling coefficient between two modes, $\gamma_{1}$ and $\gamma_{2}$ express the loss (positive number) and gain (negative number) rates of each waveguide.}
\end{figure}

In this section, we studied beam splitting properties in coupled waveguide structures based on PT symmetry, and derived the transmission matrices and their approximate forms in different phase spaces of PT-symmetry, PT-broken and EP-lines. In previous studies, transmission matrices of PT-symmetry phase have been derived \cite{Luis_1995}. However, there is almost no study on the transmission matrices and beam splitting characteristics of PT-broken phase and EP-lines. With the introduction of gain and loss into the waveguides, these modes are fundamentally changed, especially in the PT-broken case. Therefore, the transmission matrices also have new regulatory degrees of freedom, which makes the design of on-chip quantum devices more flexible.

Fig. 2 shows the schematic of two coupled waveguides, where $\mu$ expresses coupling coefficient, $E_{1}(x)$ and $E_{2}(x)$ are the evolution of electric field with transmission distance $x$. One of them is gain material with gain rate $\gamma_{1}$, while the other with loss rate $\gamma_{2}$. For the sake of derivation, we assumed that the two modes experience the same degree of gain or decay, that $\gamma \equiv -\gamma_{1}=\gamma_{2}$. In this case, the average loss $\frac{1}{2}(\gamma_{1}+\gamma_{2})$ goes to zero, then the eigenvalues change to $\lambda_{\pm}=\mp i\sqrt{\gamma^{2}-|\mu|^{2}}$ \cite{2019Exceptional}.
Depending on whether the eigenvalues are purely imaginary or complex, the eigenmodes take the form of sinusoidal or exponential functions. We start with PT-symmetry phase, corresponding to the regions between two EP-lines in the phase diagram that $|\frac{\gamma_{1}}{|\mu|} - \frac{\gamma_{2}}{|\mu|}|< 2$. The eigenvalues are real that $\lambda_{\pm}=\pm \sqrt{|\mu|^{2}-\gamma^{2}}$. The electric field intensities within two waveguides at any transmission distance are
\begin{widetext}
\begin{equation}
    \left\{\begin{array}{c}
E_{1}(x)=C_{1} e^{i \sqrt{|\mu|^{2}-\gamma^{2}} x}+C_{2} e^{-i \sqrt{|\mu|^{2}-\gamma^{2}} x} \\
E_{2}(x)=-\frac{i}{\mu}\left[\left(-\gamma+i \sqrt{|\mu|^{2}-\gamma^{2}}\right) C_{1} e^{i \sqrt{|\mu|^{2}-\gamma^{2}}x} -\left(\gamma+i \sqrt{|\mu|^{2}-\gamma^{2}}\right) C_{2} e^{-i \sqrt{|\mu|^{2}-\gamma^{2}}x} \right]
\end{array}\right. .
\end{equation}
\end{widetext}
The effect of beam splitting on the modes can be expressed in the form of transmission matrix that
\begin{equation}
    \left(\begin{array}{l}
E_{1}(x) \\
E_{2}(x)
\end{array}\right)=U(x)\left(\begin{array}{l}
E_{1}(0) \\
E_{2}(0)
\end{array}\right).
\end{equation}
$U(x)$ is transmission matrix. In PT-symmetry phase, substituting the electric field $E_{1,2}(0)$ and $E_{1,2}(x)$ satisfying Eq. (7) into Eq. (8), setting $\beta=\sqrt{|\mu|^{2}-\gamma^{2}}$, we obtained the transmission matrix $U(x)$ at the distance $x$,
\begin{widetext}
\begin{equation}
    U_{PT-symmetry}=\frac{1}{2 \beta}\left(\begin{array}{cc}
(\beta-i \gamma) e^{i \beta x}+(\beta+i \gamma) e^{-i \beta x} & \mu^{*}\left(e^{i \beta x}-e^{-i \beta x}\right) \\
\mu\left(e^{i \beta x}-e^{-i \beta x}\right) & (\beta+i \gamma) e^{i \beta x}+(\beta-i \gamma) e^{-i \beta x}
\end{array}\right) .
\end{equation}
\end{widetext}
It can be seen that the exponential terms are pure imaginary numbers, indicating two modes exchanging energy primarily. Compared with the case without gain or loss, the oscillation periods are changed. Transmission matrices in PT-phase is consistent with previous studies \cite{2019Exceptional}.

When the gain and loss of the system continue to increase and exceed the coupling coefficient that beyond the EP-lines, phase transition will occur, thus PT symmetry will be broken. This situation corresponds to the regions on either side of two EP-lines in the phase diagram, where $|\frac{\gamma_{1}}{|\mu|} - \frac{\gamma_{2}}{|\mu|}|> 2$. The oscillating modes disappear and are replaced by exponential gain or decay modes. We can get the intensities of the electric field at any position as follows,
\begin{widetext}
\begin{equation}
    \left\{\begin{array}{c}
E_{1}(x)=C_{1}^{'} e^{\sqrt{\gamma^{2}-|\mu|^{2}} x}+C_{2}^{'} e^{-\sqrt{\gamma^{2}-|\mu|^{2}} x} \\
E_{2}(x)=i\left[\frac{\gamma-\sqrt{\gamma^{2}-|\mu|^{2}}}{\mu} C_{1}^{'} e^{\sqrt{\gamma^{2}-|\mu|^{2}} x}+\frac{\gamma+\sqrt{\gamma^{2}-|\mu|^{2}}}{\mu} C_{2}^{'} e^{-\sqrt{\gamma^{2}-|\mu|^{2}} x}\right]
\end{array}\right. .
\end{equation}
\end{widetext}
It can be seen that the exponent are real numbers, and each mode contains gain and loss terms. However, the difference in coefficient results in the different degree of gain and loss experienced by the two modes. Let $\beta^{'}=\sqrt{\gamma^{2}-|\mu|^{2}}$, plugging into the transport equation Eq. (8), one can get transmission matrix $U(x)$ that
\begin{widetext}
\begin{equation}
    U_{PT-broken}=\frac{1}{2 \beta^{'}}\left(\begin{array}{cc}
(\gamma+\beta^{'}) e^{\beta^{'} x}-(\gamma-\beta^{'}) e^{-\beta^{'} x} & i \mu\left(e^{\beta^{'} x}-e^{-\beta^{'} x}\right) \\
i \mu^{*}\left(e^{\beta^{'} x}-e^{-\beta^{'} x}\right) & -(\gamma-\beta^{'}) e^{\beta^{'} x}+(\gamma+\beta^{'}) e^{-\beta^{'} x}
\end{array}\right) .
\end{equation}
\end{widetext}

For the beam splitting in PT-broken phase, the attenuation term is negligible compared to the exponential gain with sufficiently long distances. Assuming that the gain and loss coefficients $\gamma$ are sufficiently large compared with coupling coefficient $\mu$, the appointment $\sqrt{\gamma^{2}-|\mu|^{2}} \approx \gamma$ is satisfied, the transmission matrix $U(x)$ in this case can be approximated as
\begin{equation}
    U_{PT-broken}=\frac{e^{\gamma x}}{2 \gamma}\left(\begin{array}{cc}
2 \gamma & i \mu \\
i \mu^{*} & 0
\end{array}\right) .
\end{equation}
According to the transmission matrix of Eq. (12), no matter what the input state is, the light can always export from the gain end of the beam splitter, thus realizing the non-reciprocal transmission. The condition is true for all parameters, such as previously studied gain-loss \cite{2010Observation} and loss-lossless \cite{Ornigotti_2014} coupling. Hence, it also applies to gain-gain and gain-lossless coupling that not discussed in the literature. We will focus on these two cases in the next section.

Finally, the form of transmission matrix at EP-lines is given, which corresponds to the points on EP-lines that $|\frac{\gamma_{1}}{|\mu|} - \frac{\gamma_{2}}{|\mu|}|= 2$. For the specific case, $|\mu|=|\gamma|$, the eigenvalues will degenerate, i.e. $\lambda_{\pm}=\mp i\sqrt{\gamma^{2}-|\mu|^{2}}=0$. A sudden phase transition occurs. The general solutions of the differential equation are
\begin{equation}
    \left\{\begin{array}{c}
E_{1}(x)=C_{1}+C_{2} x \\
E_{2}(x)=\frac{i}{\mu}\left[|\mu| C_{1}+(|\mu| x-1) C_{2}\right]
\end{array}\right. .
\end{equation}
The transmission matrix $U(x)$ at EP-lines can then be obtained that
\begin{equation}
    U_{EP}=\left(\begin{array}{cc}
1+|\mu| x & i \mu x \\
i \mu^{*} x & 1-|\mu| x
\end{array}\right) .
\end{equation}
After verification, it can be found that both PT-symmetry phase (Eq. (9)) and PT-broken phase (Eq. (11)) can degenerate to the form of Eq. (14) when $|\gamma|_{\pm}\rightarrow|\mu|$, which verifies the self-consistency of this set of theories. So far, we have derived the transmission matrices of PT-symmetry phase, PT-broken phase and EP-lines in general coupling system. The unique transmission matrices in PT-broken phase can provide more flexible regulation of beam splitting, conducive to the design of non-reciprocal optical devices.

\begin{figure*}
\centering\includegraphics[width=12cm]{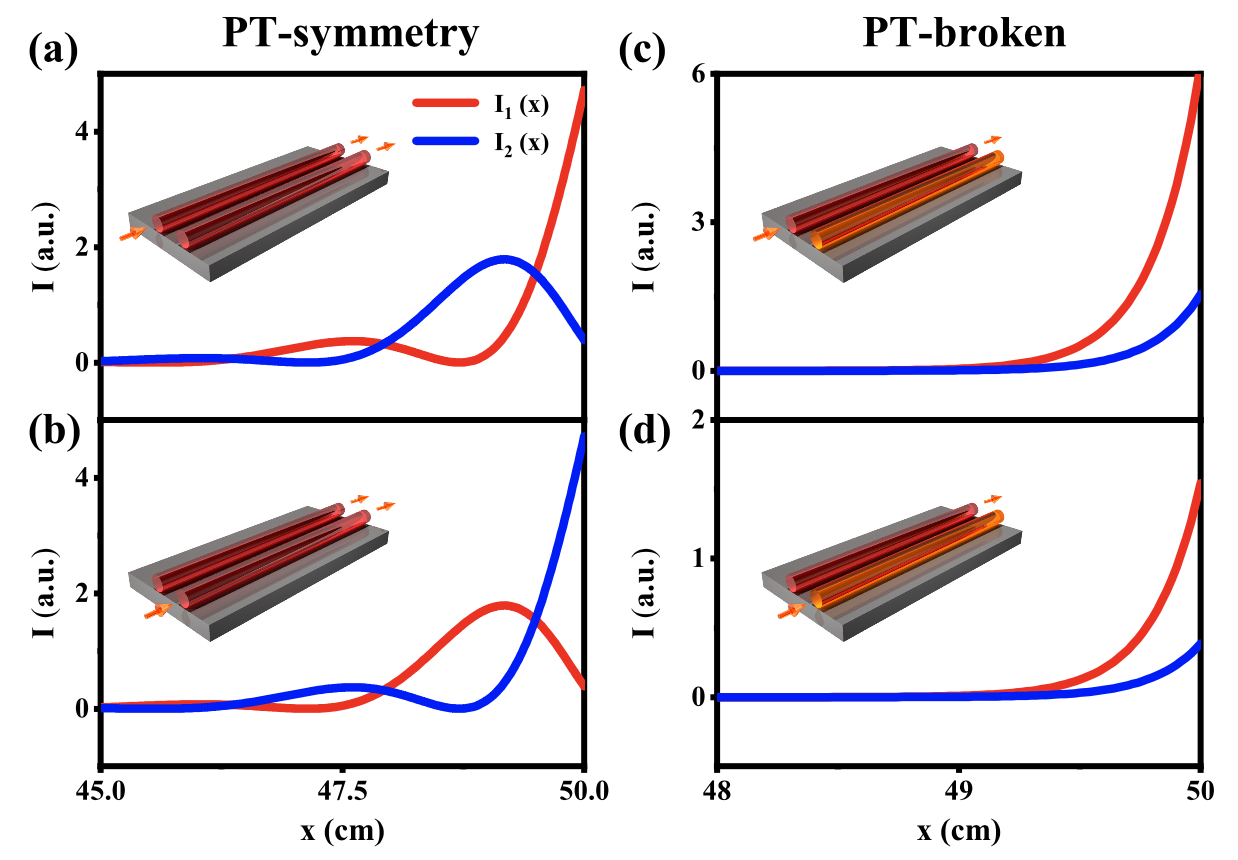}
\caption{Calculation results of gain-gain coupling. The main graph show the light intensities in two waveguides vary with the distance. Red and blue curves for $I_{1}(x)$ and $I_{2}(x)$ respectively. The subgraph is the coupled waveguide structure. The darker the cylinder, the greater the gain coefficient, and the arrows indicate the direction of input or output light.  (a) and (b) for PT-symmetry phase with $\gamma_{1}=\gamma_{2}=-0.5$cm$^{-1}$. The two modes oscillate periodically along with gain. (c) and (d) for PT-broken phase with $\gamma_{1}=-3$cm$^{-1}$, $\gamma_{2}=-0.5$cm$^{-1}$. Both modes increase exponentially and most of the energy is concentrated in the channel with higher gain.}
\end{figure*}

\section{Numerical Results}

We took coupled waveguides as an example for numerical calculation to verify PT-symmetry and broken phases. Depending on the dielectric constant of the material, individual waveguide supports exponentially enhanced or evanescent wave modes in propagation direction. When two adjacent waveguides are placed parallel, the modes will exchange energy with each other at a fixed frequency. Especially for gain-gain and gain-lossless coupling cases, the modes receive gain and exchange energy with each other as expected by phase diagram. When the gain coefficients differ greatly, most of the energy is concentrated in one mode with larger gain, showing the typical characteristics of PT-broken phase.

First, we investigated the gain-gain modes coupling that has not been paid attention to in previous studies. The mode evolution can be predicted by the phase diagram and exhibit the properties of PT-broken. This system can be used to improve the optical power of classical devices and the fidelity of quantum devices. The gain-gain coupling is expressed as red area in Fig. 1(b) of the full-parameter space phase diagram, and similarly exhibits very different properties in different phase spaces. Fig. 3 show the light intensities $I_{1}(x)$, $I_{2}(x)$ in two waveguides evolves with the propagation distance $x$, the incident light are from channel 1 ((a) and (c)) and channel 2 ((b) and (d)) respectively. The red and blue curves represent light intensity in the above and below waveguide. Here, the coupling coefficient $\mu$ is always kept at $\mu=1$cm$^{-1}$. First in PT-symmetry phase, let both waveguides have the same degree of gain that $\gamma_{1}=\gamma_{2}=-0.5$cm$^{-1}$, satisfying the condition $|\frac{\gamma_{1}}{|\mu|}-\frac{\gamma_{2}}{|\mu|}|<2$. The results are shown in Fig. 3(a) and (b). It can be seen that the waveguides experience gain as they exchange energy with each other, and the field intensity in both channels are of the same order, fitting the characteristics of PT-symmetry phase \cite{2017Non, 2018Non}. While for the PT-broken phase, let the upper waveguide go through a higher gain rate ($\gamma_{1}=-3$cm$^{-1}$) than the lower one ($\gamma_{2}=-0.5$cm$^{-1}$) and satisfy the PT-broken condition $|\frac{\gamma_{1}}{|\mu|}-\frac{\gamma_{2}}{|\mu|}|>2$. From Fig. 3(c) and (d) we can see, regardless of which port the light comes in from, the field intensity of the top channel is about 3 times stronger than that of the bottom over a certain distance, showing most of the energy is concentrated in the waveguide with higher gain, fitting the characteristics of PT-broken phase \cite{2017Non, 2018Non}. Therefore, the numerical results of gain-gain coupling are consistent with the theoretical expectation.

Then we focused on the gain-lossless modes coupling, corresponding to the negative half coordinate axis of the phase diagram. Let $\gamma_{1}=-0.5$cm$^{-1}$, $\gamma_{2}=0$ and $\mu=1$cm$^{-1}$, which satisfies the condition of PT-symmetry that $|\frac{\gamma_{1}}{|\mu|}-\frac{\gamma_{2}}{|\mu|}|<2$. Fig. 4(a) and (b) show the calculation results of light from the upper and lower waveguides respectively. The oscillatory transmission modes accompanying gain are found in both channels. The light of two output ends have similar intensity. Then, increase the gain coefficient and let $\gamma_{1}=-3$cm$^{-1}$, satisfying the condition of PT-broken that $|\frac{\gamma_{1}}{|\mu|}-\frac{\gamma_{2}}{|\mu|}|>2$. When the light is incident from the gain waveguide (shown in Fig. 4(c)), the electric field strength of both channels increase exponentially. While when the incident light is from the lossless channel (shown in Fig. 4(d)), they first go though a prominent energy exchange and then grow exponentially. From the results in subgraphs of Fig. 4(c) and (d), whether the light entered from the gain port or the lossless port, most of the energy is concentrated in the gain waveguide after coupling. The above characteristics also fit the predictions.

\begin{figure*}
\centering\includegraphics[width=12cm]{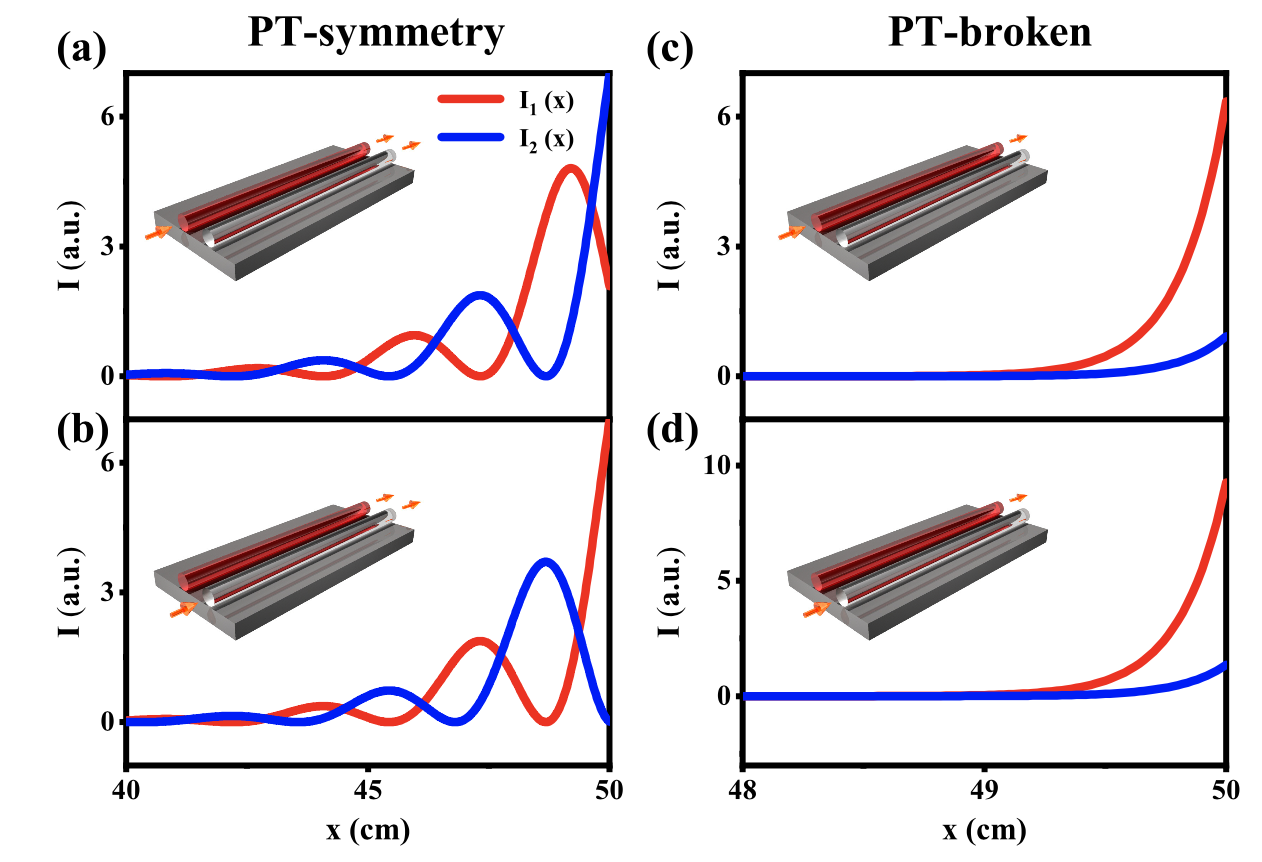}
\caption{Calculation results of gain-lossless coupling. The main graph show the light intensities in two waveguides vary with the distance. Red and blue curves for $I_{1}(x)$ and $I_{2}(x)$ respectively. The subgraph is the coupled waveguide structure. Red cylinder for gain medium, and white cylinder for lossless material. The arrows indicate the direction of input or output light. (a) and (b) for PT-symmetry phase with $\gamma_{1}=-0.5$cm$^{-1}$, $\gamma_{2}=0$. The two modes oscillate periodically along with gain. (c) and (d) for PT-broken phase with $\gamma_{1}=-3$cm$^{-1}$, $\gamma_{2}=0$. Both modes increase exponentially and most of the energy is concentrated in the gain channel.}
\end{figure*}

Finally, we calculated the loss-loss and gain-loss coupling, which corresponds to the grey area in the phase diagram. The above two systems have been widely discussed in previous studies. For gain-loss coupling, both modes show the same degree of periodic oscillation in PT-symmetry phase. While in PT-broken phase, they become exponential growth modes and the gain waveguide dominate. For two loss modes, they exhibit periodic exchange of energy with losses in PT-symmetry phase and exponential in PT-broken phase. By comparing the two situations, the above conclusions are not affected by the change of incident light, which are consistent with the results of previous studies \cite{2017Non, 2018Non}.

In this part, we calculated the light intensity of coupled waveguides, involving all quadrants and coordinate axes of the PT phase diagram. It can be seen that the results are consistent with the theoretical analysis. In particular, we obtained the mode evolution in the gain-gain and gain-lossless coupling that almost have not been discussed, which combines the unique modes of PT symmetry with higher power or higher fidelity of gain medium, thus expanding the design ideas of classical or quantum optical devices.

\section{Conclusion}
We have demonstrated the phase diagram of PT symmetry system in a full-parameter space containing gain, lossless and loss. The phase diagram has been divided into PT-symmetry and PT-broken phases by EP-lines, which can be used to predict the phase of coupled micro/nano photonic structures. Then, we have conducted numerical calculation with coupled waveguides as example, paying special attention to gain-gain and gain-lossless coupling. We have also discussed the beam splitting in non-Hermitian systems, and have deduced the transmission matrices in different phase. In PT-symmetry phase, oscillatory transport dominates, whereas PT-broken phase corresponds to exponential gain or decay modes. Therefore, the proposed PT symmetry provides a new manipulation degree of freedom for beam splitting. Investigating PT symmetry has grown exponentially in recent years, and is still a bright future for new insights. 

Emerging photonic technologies provide wider playground for studying non-Hermitian quantum mechanics, including phase transitions, conservation relations and spontaneous symmetry. Not limited to basic physics, joining PT symmetry with micro/nano photonics might also dramatically improve the performance and robustness of on-chip devices.

\begin{acknowledgments}
 National Natural Science Foundation of China (11974032, 11734001, 11525414); Key R\&D Program of Guangdong Province (2018B030329001).
\end{acknowledgments}

\nocite{*}


\end{document}